# Enhancing resonant second harmonic generation in bilayer WSe$_2$ by layer-dependent exciton-polaron effect

Soonyoung Cha[1,†], Tianyi Ouyang[1,†], Takashi Taniguchi[2], Kenji Watanabe[3], Nathaniel M. Gabor[1*] & Chun Hung Lui[1*]

[1] Department of Physics and Astronomy, University of California, Riverside, CA, USA.

[2] International Center for Materials Nanoarchitectonics, National Institute for Materials Science, Tsukuba, Japan.

[3] Research Center for Functional Materials, National Institute for Materials Science, Tsukuba, Japan.

[†]These authors contributed equally to this work

[*]Corresponding authors. Email: nathaniel.gabor@ucr.edu; joshua.lui@ucr.edu

Two-dimensional (2D) materials serve as exceptional platforms for controlled second harmonic generation (SHG), an important nonlinear optical phenomenon with diverse applications. Current approaches to SHG control often depend on non-resonant conditions or symmetry breaking via single-gate control. Here, we employ dual-gate bilayer WSe$_2$ to demonstrate a new SHG enhancement concept that leverages strong exciton resonance and layer-dependent exciton-polaron effect. By selectively localizing injected holes within one layer, we induce exciton-polaron states in the hole-filled layer while maintaining normal exciton states in the charge-neutral layer. The distinct resonant conditions of these layers effectively break interlayer inversion symmetry, thereby promoting resonant SHG. Our method achieves a remarkable 40-fold enhancement of SHG at minimal electric field, equivalent to conditions near the dielectric-breakdown threshold but using only ~3% of the critical breakdown field. Our findings also reveal significant sensitivity of resonant SHG to carrier density and carrier type, with distinct enhancement and quenching observed across different gating regimes. This advancement offers an innovative approach to manipulating SHG in 2D excitonic materials and provides a potent spectroscopic tool for probing layer-dependent quantum states.

Second harmonic generation (SHG), where two identical photons combine into a single photon with double frequency, plays a pivotal role in various scientific disciplines such as nonlinear frequency conversion, quantum optics, materials characterization, surface studies, and biological imaging [1-3]. Two-dimensional (2D) crystals, including transition metal dichalcogenides (TMDs) like $MoS_2$ and $WSe_2$, stand out as exceptional media for SHG. These materials exhibit selective inversion symmetry based on their layer numbers [4,5] and stacking arrangements [6-9]. For instance, monolayer $WSe_2$, lacking inversion symmetry, exhibits robust SHG [10,11]. However, stacking two such monolayers with opposite layer orientations results in SHG phases that cancel each other out, leading to no net SHG (Fig. 1a).

In addition to varying layer thickness, researchers explore different approaches to control SHG in 2D crystals [12-15]. Notably, applying an electric field can break the inversion symmetry of bilayers, thereby disrupting destructive SHG interference between layers and promoting emergent SHG (Fig. 1b). However, the SHG efficiency remains relatively low due to the weak modulation of quantum states by the electric field. Moreover, increasing the electric field poses risks such as dielectric breakdown and instabilities in devices. Other methods, such as surface layer attachment [14] and layer-dependent phase transitions [15], exhibit inconvenience and limited effectiveness.

In this study, we focus on bilayer $WSe_2$ as a model system to demonstrate an innovative approach to enhance SHG. Our method leverages the strong exciton resonance and the distinctive layer-dependent exciton-polaron effect in 2D TMD semiconductors. Specifically, we exploit two key features of typical TMD bilayers: (1) their holes are localized within one layer; (2) they host robust intralayer excitons that interact strongly with ambient holes in the same layer to form layer-specific exciton polarons. By applying an electric field to create an interlayer potential difference, we selectively inject holes into one layer while maintaining the other layer charge-neutral. This approach allows us to create normal exciton states in the charge-neutral layer but exciton-polaron states in the hole-filled layer. Unlike excitons, exciton polarons exhibit lower energy and much weaker resonant effect with SHG [11]. When performing SHG at the exciton energy, the layer with exciton states generates strong resonant second harmonic waves, while the layer with exciton-polaron states exhibits minimal response. The lack of destructive interference between them results in a pronounced net SHG signal (Fig. 1c).

Our approach, founded on the layer-dependent exciton-polaron effect, represents a conceptual departure from previous methods that depend on non-resonant conditions and/or symmetry breaking via single-gate control [16-18]. By independently controlling charge density and electric field in dual-gate devices, our method demonstrates a remarkable 40-fold enhancement of SHG at minimal electric field. Notably, we achieve SHG emission comparable to that induced near the dielectric-breakdown threshold with just ~3% of the critical breakdown field. Achieving such substantial SHG enhancement at minimal electric fields opens new avenues for experimental setups and applications where high electric fields are impractical. Moreover, our findings underscore a remarkable sensitivity of SHG to variation in carrier density and carrier type (electron or hole), showcasing different enhancement and quenching behaviors across different gating regimes. This nuanced understanding paves the way for tailored manipulation of SHG in 2D materials.

**Results and Discussion**

We fabricate dual-gated 2H-stacked bilayer $WSe_2$ encapsulated by hexagonal boron nitride (BN). Thin graphite flakes are used as contacts and gate electrodes to improve device performance. The dual-gate configuration enables us to independently control the charge density (electric field) by applying voltages of the same sign (opposite signs) to the top and bottom gates. We measure the SHG from bilayer $WSe_2$ under the excitation of an ultrafast laser with tunable wavelength (PHAROS + ORPHEUS, Light Conversion Inc.). All measurements are conducted at sample temperature $T \approx 6$ K. Detailed experimental methods are provided in the Methods section. The results in the main paper are obtained on Device 1, and additional results on Device 2 are presented in the Supporting Information.

**SHG at varying electric field and charge density.** We first compare the influences of electric field and charge density on the SHG of bilayer $WSe_2$ (Figure 2). In the measurement, we set the laser wavelength to be 1449 nm so that the second harmonic energy (1.712 eV) matches the A-exciton resonance in bilayer $WSe_2$. The A exciton, consisting of an electron-hole pair in the K valley, is essentially an intralayer state because both the electron and hole wavefunctions are predominantly localized in the same layer [19,20]. Indeed, we can consider the bilayer K valleys as two decoupled monolayer valleys in Layer One and Layer Two (denoted as L1 and L2 valleys), whose energies are the same because the two layers are connected by inversion symmetry. When an electric field is applied to break the inversion symmetry, the degeneracy between the L1 and L2 valleys will be lifted (Figure 2a).

Figure 2b-c displays the SHG under varying electric field ($E$) at zero carrier density ($N$). In these conditions, both layers are charge-neutral (denoted by *i-i* regime since both layers are intrinsic). At zero field, the SHG signal is zero as expected because of inversion symmetry. But when the field increases, the SHG emerges due to the breaking of inversion symmetry [17]. The electric field can reach $E \sim 0.15$ V/nm before the device undergoes dielectric breakdown. In comparison, we also measure the bilayer SHG at varying charge density but zero electric field (Figure 2d-f). As inversion symmetry is preserved in this case, the two monolayer valleys remain energy-degenerate. The injected charges are distributed

equally between two layers (Figure 2d). Consequently, no appreciable SHG is observed for both electron and hole injection (Figure 2e-f).

Intriguing SHG behavior emerges when we combine both electric field and charge injection (Figure 2g-i). In this scenario, the two monolayer K valleys have different energies because of the electric field. When holes are injected into the sample, they reside exclusively in one layer (Figure 2g). This electronic layer polarization augments their asymmetry, leading to an enhancement of SHG. This can be clearly seen in Figure 2h-i, which display the SHG map and intensity at varying charge density under a moderate fixed electric field ($E$ = 0.024 V/nm). Here $E$ represents the calculated electric field without considering the screening effect of the injected carriers [21,22] (see the Supporting Information for the calculation of interlayer electric field with and without considering the free-carrier screening effect).

We can interpret the second harmonic profile in Figure 2i across four gating regimes (denoted by the charge density types *n, i-i, i-p, p-p*) as designated in Figure 2g, i. When the Fermi level resides within the band gap, bilayer $WSe_2$ is in the *i-i* regime as both L1 and L2 are intrinsic. Here, the SHG is already observable due to the electric field, just like the SHG in Figure 2c. Upon injecting holes into bilayer $WSe_2$, it goes into the *i-p* regime, where L1 remains intrinsic but L2 becomes populated with holes. Remarkably, the SHG is much enhanced in this regime. In particular, at hole density ~ $4 \times 10^{12}$ cm$^{-2}$, it becomes five times stronger than in the *i-i* regime. This heightened SHG intensity surpasses the maximum achievable SHG (denoted by a star in Figure 2c) solely induced by an electric field without charge, limited by the breakdown of gate dielectrics. Therefore, by employing charge injection, we can surpass the dielectric-breakdown limit of field-induced SHG by using an electric field (0.024 V/nm) well below the breakdown critical field (~0.15 V/nm).

However, the SHG is quenched when bilayer $WSe_2$ transitions into the *p-p* regime, where both L1 and L2 are populated with holes (Figure 2g-i). After the Fermi level reaches the K valley at the other layer (marked by a square in Figure 2g), any additional injected holes distribute equally between the two layers. These extra holes do not contribute to SHG, but potentially quench it. On the other hand, the SHG is also quenched when we inject electrons to reach the *n* regime (Figure 2i). This is because the conduction band minimum in bilayer $WSe_2$ resides in the Q valley (marked by a circle in Figure 2g), not the K valley. Unlike the localized K-valley states, the Q-valley states are not confined to one layer [19]. Consequently, the accumulated electrons in the Q valleys cannot efficiently break the inversion symmetry to promote SHG. In summary, the SHG is enhanced exclusively in the *i-p* regime but quenched in the *p-p* and *n* regime. Such a sensitive dependence on carrier density and type is uncommon in SHG phenomena.

**Comparison between SHG and reflectance contrast.** To explore the mechanism behind the SHG enhancement, we investigate its correlation with excitonic resonances in bilayer

WSe$_2$. To this end, we measure the reflectance contrast ($\Delta R/R$) of our bilayer WSe$_2$ sample and calculate the second-order energy derivative $d^2(\Delta R/R)/dE^2$ to reveal subtle features. Figure 3a displays a color map of $d^2(\Delta R/R)/dE^2$ across varying charge densities under a fixed electric field $E = 0.024$ V/nm. The map shows abrupt changes in excitonic features at critical charge densities (indicated by open symbols), which are attributed to Fermi levels reaching the conduction band minimum (circle), valence band maximum (triangle), and the opposing monolayer's valence band (square), thus defining four distinct regimes: *n, i-i, i-p, p-p* (Figure 2g, 3a).

The *n* regime exhibits optical features from coupled states between K-valley excitons and Q-valley Fermi seas. The *i-i* regime exhibits an optical feature from two degenerate intralayer excitons ($A_{L1}^0$, $A_{L2}^0$) (Figure 3b). Transitioning to the *i-p* regime reveals two optical features: a strong, higher-energy one attributed to the $A_{L1}^0$ intralayer exciton and a weak, lower-energy one attributed to the $A_{L2}^+$ intralayer exciton polaron (Figure 3a,c) [23-25]. The *p-p* regime shows two weak optical features, which are attributed to the $A_{L1}^+$, $A_{L2}^+$ intralayer exciton polarons (Figure 3a,d). We neglect the relatively weak interlayer coupling [26].

To further exlore different gating regimes, we tuned the laser wavelength to sweep the SHG energy across the excitonic spectral range. Figure 3a overlays SHG intensities (solid dots) as a function of sweeping two-photon energy in the *i-i, i-p, p-p* regimes. The SHG profiles exhibit two-photon resonant features that correspond closely to excitonic resonances in these regimes [11,27,28]. Particularly in the *i-p* regime where SHG peaks, a pronounced resonance is observed with the bright $A_{L1}^0$ state, contrasting with minimal resonance with the dim $A_{L2}^+$ state (Figure 3a). This distinction suggests that L1 generates strong second-harmonic signals while L2 generates weak ones, resulting in a lack of destructive interference between them as depicted in Figure 1c.

While Figure 3 shows data at a specific electric field ($E = 0.024$ V/nm), systematic adjustment of the electric field modulates the gating regimes. Figure 4a-e present five maps of $d^2(\Delta R/R)/dE^2$ at $E = 0, \pm 0.024, \pm 0.048$ V/nm alongside valley configurations (Additional results can be found in the Supporting Information). The *n* and *i-i* regimes show minimal variation across electric fields, while the *i-p* regime expands nearly linearly with increasing field strength because the opposing Stark shifts of L1 and L2 valleys widens their energy separation. In comparison, the field- and density-dependent SHG map (Figure 4f) also shows four distinct regimes, aligning roughly with the *n, i-i, i-p, p-p* regimes observed in the reflection measurements (highlighted by lines in Figure 4f). Notably, the SHG is pronounced in the *i-p* regime, with the density range expanding with increasing electric field strength.

The comparison of reflection and SHG data provides crucial insights into the mechanism of SHG enhancement. In the *i-i* regime, both layers host intralayer exciton states with equal SHG resonant effect, leading to destructive interference and zero net SHG (Figure 1a). Conversely, in the *i-p* regime, the *i*-layer with exciton states and the *p*-layer with exciton polaron states exhibit distinct resonant behaviors. Prior research showed that exciton polarons have much weaker resonant effect with SHG than normal excitons [11]. When performing SHG at the excitonic resonance, the *i*-layer emits strong resonant second harmonic signals while the *p*-layer produces weak ones. The lack of destructive interference between them results in pronounced net SHG (Figure 1c). In the *p-p* regime, both layers host exciton polaron states with weak resonant effect, leading to overall weak SHG. Similarly, in the *n*-type regime where the Q-valley electrons distribute across both layers, the exciton polaron states form across both layers, resulting in weak SHG.

**SHG enhancement in the weak-field regime.** We observe a striking phenomenon upon scrutinizing the weak-field region in Figure 4f. Here the field, exemplified by $E = 0.005$ V/nm (indicated by the orange dashed line), accounts for only ~3% of the critical field $E \sim 0.15$ V/nm as denoted by the star in Figure 2c near dielectric breakdown. At this field strength, SHG remains negligible in the charge-neutrality regime, indicaitng that the field strength is insufficient to stimulate SHG. However, within this field range, the SHG experiences a sudden enhancement to a robust intensity within a narrow band of hole density around $3.1 \times 10^{12}$ cm$^{-2}$, as depicted by the cross-cut orange profile in Figure 4g. This SHG intensity closely rivals the maximum achievable SHG at $E = 0.15$ V/nm (near dielectric breakdown) with no charge injection, as marked by the star in Figure 2c and 4h. Essentially, by precisely injecting charges into bilayer WSe$_2$, we can achieve highly efficient SHG, comparable to the SHG near dielectric breakdown with no charge, with a mere ~3% of the breakdown critical field. This underscores an exceptionally sensitive and efficient method for controlling SHG.

To quantify the enhanced efficiency of SHG, Figure 4h presents a map of the enhancement ratio across various charge densities (y-axis) relative to the SHG intensity at zero charge density under the same electric field (x-axis). Across most of the *i-p* regime, the enhancement ratio ranges between 3 and 5, as illustrated by the black cross-cut profile in Figure 4i. However, within narrow ranges near charge density $N = 4 \times 10^{12}$ cm$^{-2}$ and electric fields $E = \pm 0.01$ V/nm, the enhancement ratio peaks at about 40, as indicated by the red cross-cut profile in Figure 4i. This underscores the remarkable efficiency enhancement achieved through our strategy.

In conclusion, we demonstrate a novel approach to enhance SHG in bilayer WSe$_2$ by leveraging the strong excitonic resonance and layer-dependent exciton-polaron effect. Our findings also reveal a remarkable sensitivity of SHG to carrier density and carrier type, with distinct enhancement and quenching observed in different gating regimes. This detailed understanding paves the way for advanced manipulation of SHG in bilayer WSe$_2$ and potentially some other 2D materials. Given the intimate relationship between SHG and intralayer excitonic response, SHG holds promise as a sensitive probe for exploring layer-dependent quantum states in 2D materials [29,30]. Furthermore, our analysis on the critical points of the *i-p* to *p-p* transition suggests that many-body interactions play a crucial role on the electrostatic properties of bilayer WSe$_2$ (see Section 5 in the Supporting Information) [31-33]. Such interactions may be related to the U-shaped contour observed in the SHG map at low carrier densities (Fig. 4f) and the sharp SHG enhancement at low electric fields (Fig. 4h), which are not anticipated from the reflectance contrast data (open symbols in Fig. 4f). Further research is merited to explore the potential of using SHG as a sensitive probe for investigating many-body interactions in 2D materials.

## Methods

**Device fabrication.** We fabricate dual-gate bilayer WSe$_2$ devices encapsulated by hexagonal boron nitride (BN) on silicon substrate with a 285-nm SiO$_2$ epilayer. Initially, we employ standard electron beam lithography and electron beam evaporation techniques to deposit 5-nm-thick titanium/50-nm-thick gold electrodes onto SiO$_2$/Si substrates. Next, thin graphite flakes, BN flakes, and bilayer WSe$_2$ flakes are exfoliated from their bulk crystals onto SiO$_2$/Si substrates. Our bulk WSe$_2$ crystals are procured from HQ Graphene Incorporation. Subsequently, a dry transfer method is utilized to stack these different thin crystal flakes together. Sequentially, we use polycarbonate stamps to pick up a thin graphite flake (serving as the top-gate electrode), a BN flake (top-gate dielectric), bilayer WSe$_2$, second graphite flake (contact electrode), another BN flake (bottom-gate dielectric), and third graphite flake (bottom-gate electrode). Finally, the stack of materials is transferred onto a pre-patterned SiO$_2$/Si substrate, ensuring proper alignment of the graphite electrodes with the Ti/Au pre-patterned electrodes on the substrate.

**Reflectance contrast measurement.** The reflection contrast measurements are conducted using an optical cryostat (Montana Instruments) with estimated sample temperature of approximately 6 K. A broadband light source from Thorlabs (SLS201L) is employed for these measurements. The white light is focused onto the sample by a microscope objective with a numerical aperture of 0.6, resulting in a spot diameter of ~2 μm on the sample surface. Subsequently, the reflected light is collected by the same objective and directed towards a high-resolution spectrometer (Princeton Instruments) equipped with a charge-coupled-device (CCD) camera (Princeton Instruments, PIXIS 400) for analysis. To obtain a reflectance contrast spectrum of bilayer WSe$_2$, two spectra are measured: one from the sample area containing bilayer WSe$_2$ (referred to as $R_s$), and another from a nearby reference area lacking bilayer WSe$_2$ (referred to as $R_r$, specifically the Gr/BN/BN/Gr area). The reflectance contrast is calculated using the formula $\Delta R/R = (R_s - R_r)/R_r$. To enhance the visibility of weak features, we compute the second-order energy derivative of the $\Delta R/R$ spectra.

**Second-harmonic generation measurement.** In our second-harmonic generation (SHG) experiment, we utilize an ultrafast laser system (PHAROS + ORPHEUS) provided by Light Conversion. The light source comes from a mode-locked Yb: KGW laser (PHAROS) with a centrol emission wavelength of 1030 nm and a repetition rate of 250 kHz. Through an optical parametric amplifier (ORPHEUS), the laser wavelength is converted to the infrared range (1400 ~ 1500 nm). The infrared beam is focused onto the sample using a microscope objective (NA = 0.6). The spot diameter is ~2 µm on the sample. The generated second harmonic signal is collected by the same objective, then filtered through an 800-nm short-pass filter (Thorlabs, FESH0800) to block the excitation beam. Subsequently, the signal is analyzed by a high-resolution spectrometer (Princeton Instruments) equipped with a charge-coupled-device (CCD) camera (Princeton Instruments, PIXIS 400). Linear polarizers and half-wave plates are inserted into the beam path to control the polarization of the excitation laser and measure the polarization of the detected light.

**Utilization of ChatGPT.** The manuscript was prepared with the assistance of ChatGPT, a large language model developed by OpenAI. ChatGPT aids in generating and refining the manuscript text. Nevertheless, accountability for the work remains with the human authors, with ChatGPT serving as a tool in the writing process.

**Acknowledgements**
C.H.L. acknowledges support from the National Science Foundation (NSF) Division of Materials Research CAREER Award No. 1945660 and from the American Chemical Society Petroleum Research Fund No. 61640-ND6. K.W. and T.T. acknowledge support from the JSPS KAKENHI (Grant Numbers 19H05790, 20H00354 and 21H05233). N.M.G. and S.C. were supported by the Army Research Office Electronics Division Award no.W911NF2110260, and the Presidential Early Career Award for Scientists and Engineers (PECASE) through the Air Force Office of Scientific Research (award no. FA9550-20-1-0097).

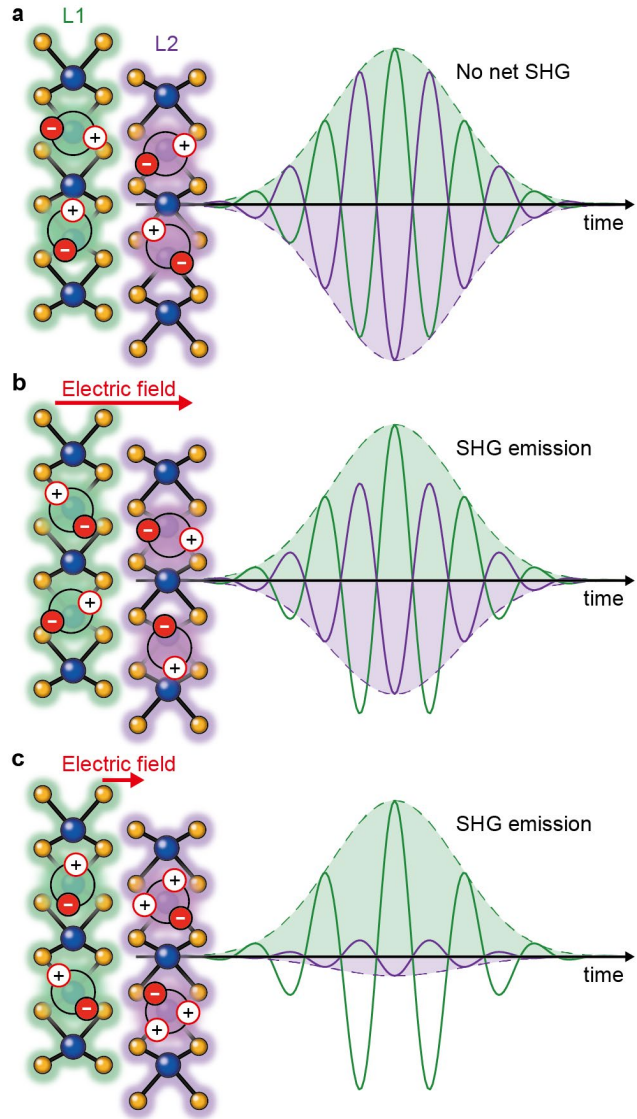

**Figure 1 | Schematic resonant second harmonic generation (SHG) from 2H-stacked bilayer WSe$_2$. a,** Both WSe$_2$ layers (L1 and L2) host intralayer exciton states, producing opposite second harmonic signals that cancel each other out, resulting in zero net SHG. **b,** Application of an inversion-symmetry-breaking electric field disrupts the destructive interference, allowing for a finite net SHG signal. **c,** By selectively injecting holes into one layer to induce layer-specific exciton-polaron states, while maintaining normal exciton states the charge-neutral layer, distinct resonant behaviors of the two layers lead to enhanced SHG signals.

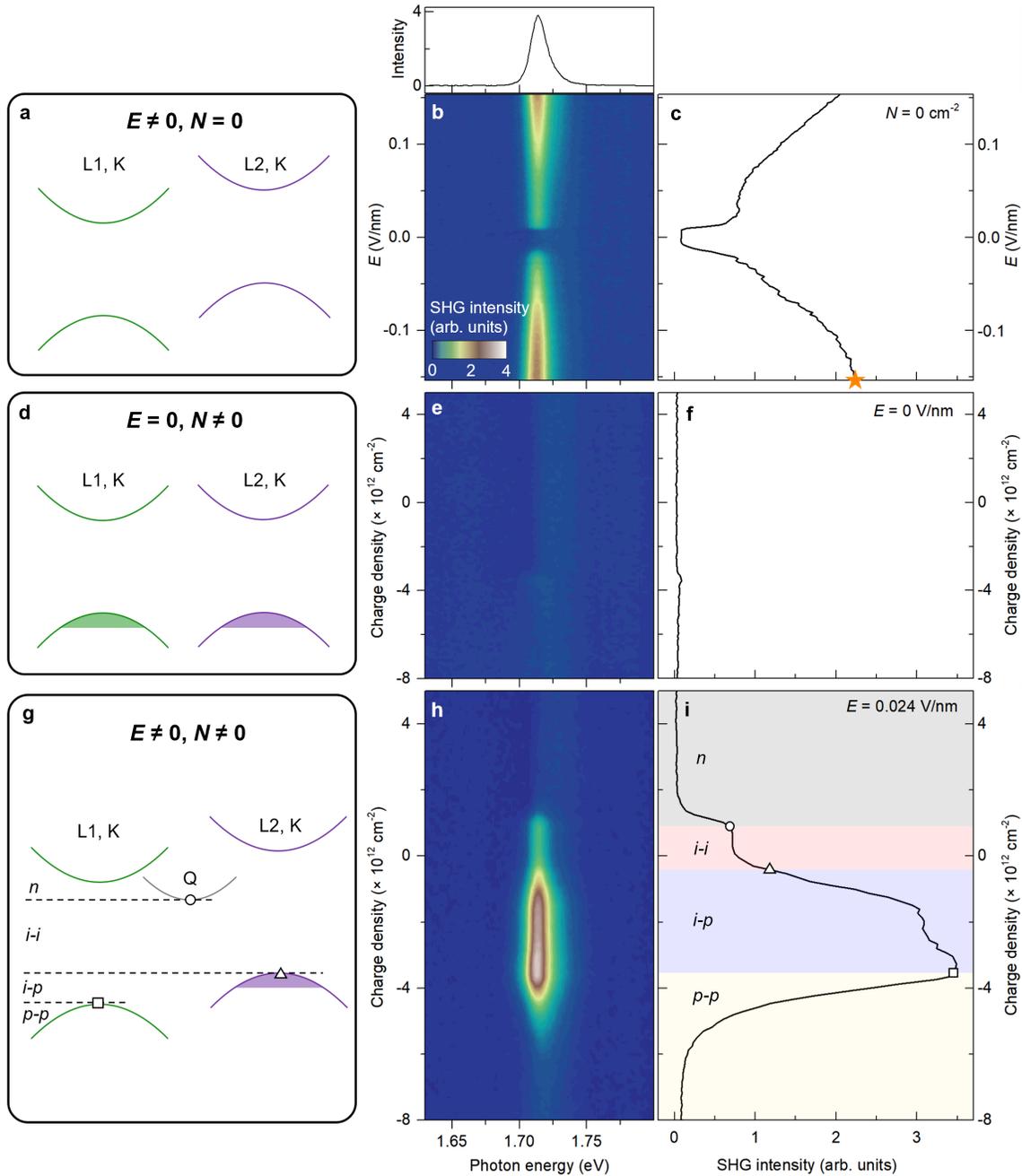

**Figure 2 | Influences of electric field and charge density on the second harmonic generation (SHG) of bilayer WSe$_2$. a,** Schematic band structure of the K valleys in Layer One (L1) and Layer Two (L2). The application of an out-of-plane electric field $E$ lifts the layer valley degeneracy. **b-c,** Second harmonic spectral map and integrated intensity at varying electric field and zero charge density. Top panel of b displays a representative SHG spectrum. The star at panel c denotes the SHG intensity near dielectric breakdown. **d-f,** Similar plots to panels a-c for zero electric field but varying charge density. **g-i,** Similar plots for a finite electric field ($E$ = 0.024 V/nm) and varying charge density. In panel g, the

Q valley is added for clarity. The dashed lines with circle, triangle, and square denote different critical Fermi levels that define the *n, i-i, i-p, p-p* charging regimes of the two layers. These different regimes with corresponding separation symbols are also denoted in panel i.

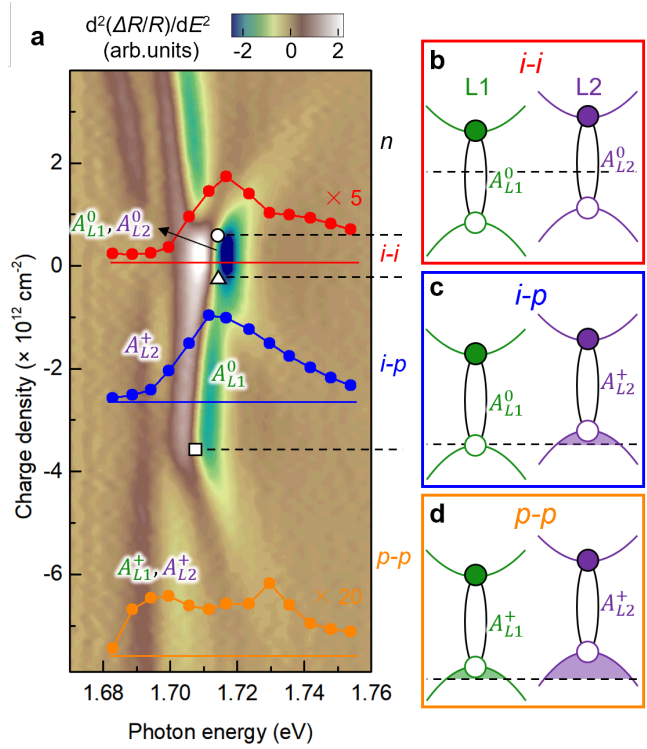

**Figure 3 | Comparing second harmonic generation (SHG) and excitonic resonances in bilayer WSe₂. a,** Charge-density-dependent color map of the second-order energy derivative of reflectance contrast $[d^2(\Delta R/R)/dE^2]$ for bilayer WSe2 under an out-of-plane electric field $E$ = 0.024 V/nm. The red, blue, orange lines denote three selected charge densities. The solid dots of corresponding color represent the integrated SHG intensity as a function of double laser photon energy at these three densities. The red and orange SHG signals are magnified for 5 and 20 times, respectively. **b,** Band diagram at the red line in panel a. Here both layers (L1, L2) are intrinsic (*i-i* regime), giving rise to two degenerate intralayer excitons ($A^0_{L1}$, $A^0_{L2}$). **c,** Band diagram at the blue line in panel a, where L1 is intrinsic but L2 is populated with holes (*i-p* regime), giving rise to the $A^0_{L1}$ exciton and $A^+_{L2}$ intralayer exciton polaron. **c,** Band diagram at the orange line in panel a. Here both layers are populated with holes (*p-p* regime), giving rise to intralayer exciton polarons on both layers ($A^+_{L1}$, $A^+_{L2}$).

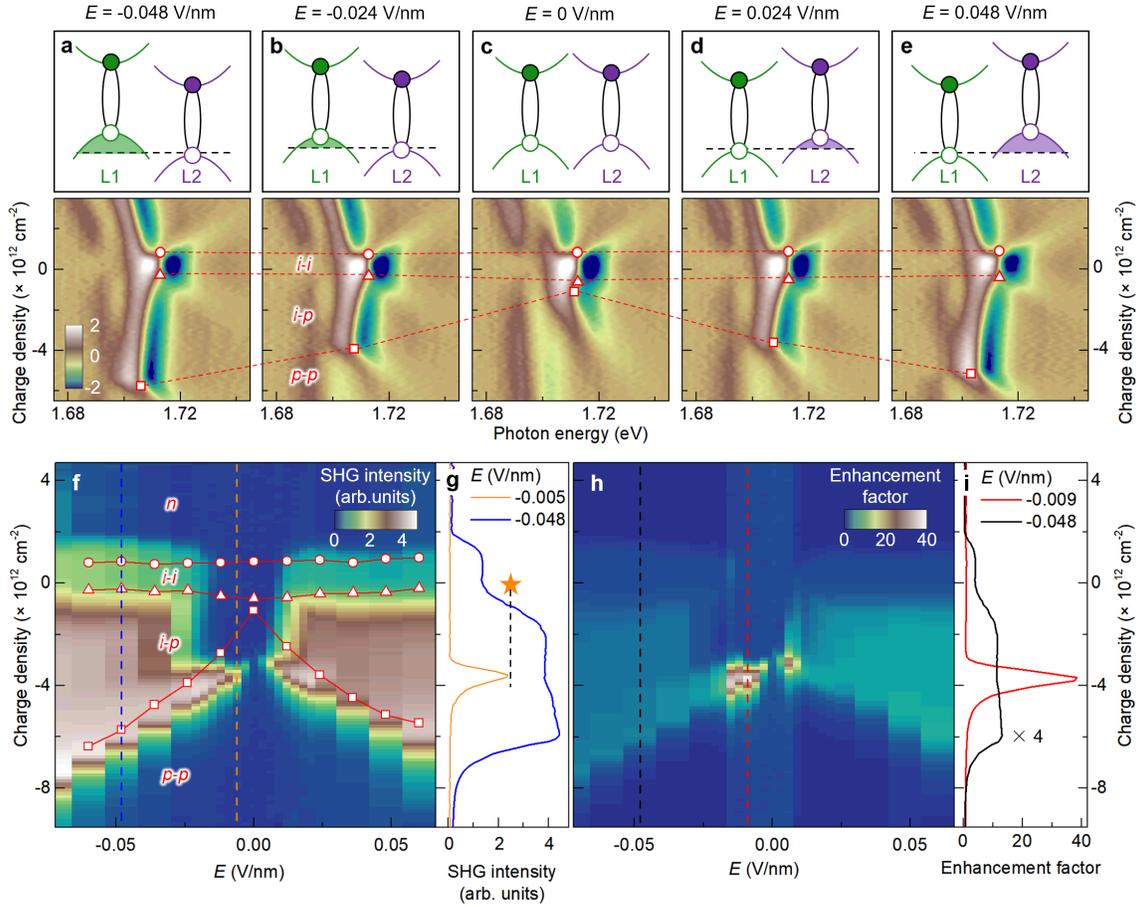

**Figure 4 | Electric field and charge density modulation of SHG regimes. a-e,** Second-order differential reflectance contrast maps of bilayer $WSe_2$ along with the corresponding K-valley configurations of the two layers at five different electric fields $E = 0, \pm 0.024, \pm 0.048$ V/nm. Like in Figure 2g, the circles, triangles, squares denote the critial charge densities where the Fermi level reaches the conduction band, the first valence band, and the second valence valley, respectively. The charge density regions between them defines the $n$, $i$-$i$, $i$-$p$, $p$-$p$ regimes, as delineated by the dashed lines. **f,** The SHG intensity color map as a function of charge density and electric field, measured at laser wavelength of 1449 nm (two-photon energy 1.712 eV). The circles, triangles, and squares mark the critical charge densities obtained from the reflection data, from which the $n$, $i$-$i$, $i$-$p$, $p$-$p$ regimes are deduced. **g,** The charge-density-dependent SHG intensity profile at $E$ = -0.005 and -0.048 V/nm. The star denotes the SHG intensity at charge neutrality and $E$ = -0.15 V/nm (near dielectric breakdown) as marked in Figure 2c. **h,** The SHG enhancement ratio, extracted from panel f, relative to the SHG intensity at zero charge density under the same electric field. **i,** Two cross-cut profiles along the red and black dashed lines ($E$ = -0.009 and -0.048 V/nm) in panel h. The black profile is magnified by 4 times for clarify.

14